# Magnons and crystal field transitions in hexagonal $RE$MnO$_3$ ($RE$ = Er, Tm, Yb, Lu) single crystals

E. C. Standard, T. Stanislavchuk, and A. A. Sirenko[1]

*Department of Physics, New Jersey Institute of Technology, Newark, New Jersey 07102, USA*

N. Lee and S.-W. Cheong

*Rutgers Center for Emergent Materials and Department of Physics and Astronomy, Rutgers University, Piscataway, New Jersey 08854, USA*

## ABSTRACT

Far-infrared optical transmission spectra of the antiferromagnetic resonances, or magnons, and crystal field transitions have been studied in hexagonal $RE$MnO$_3$ ($RE$ = Er, Tm, Yb, Lu) single crystals. The magnon and CF frequencies, their oscillator strengths, and effective $g$-factors have been measured using external magnetic fields up to 9 T in the temperature range between 1.5 K and 100 K. The magnon frequency increases systematically with a decrease of the $RE$ ion radius. The magnetic ordering of $RE$ ions ($RE$=Er, Tm, Yb) was observed at low temperatures $T$<3.5 K and in strong external magnetic fields. The observed effects are analyzed taking into account main magnetic interactions in the system including exchange of the Mn$^{3+}$ spins with $RE^{3+}$ paramagnetic moments.

---

[1] sirenko@njit.edu

I. INTRODUCTION

Rare-earth manganites $RE$MnO$_3$ attracted recently a lot of attention due to their intriguing structural, magnetic, and multiferroic properties. The compounds, grown at ambient pressure, usually have either orthorhombic ($RE$=La, ... Dy) or hexagonal structure ($RE$= Ho, …Lu, and Sc, Y), where the choice of the structure is determined by the $RE^{3+}$ ionic radius $r_i$.[1,2] Recently, multiferroic effects, such as the coupling between the ferroelectric and magnetic orders, have been found in both hexagonal and orthorhombic $RE$MnO$_3$ compounds, as well as in materials with both non- magnetic $RE$ ions, such as in YMnO$_3$ and LuMnO$_3$, and magnetic $RE$ ions with the incomplete 4$f$ shell.[3,4,5,6] The difference between the major mechanisms of multiferroicity in orthorhombic and hexagonal manganites is in the focus of modern theoretical and experimental work.[7,8,9,10,11,12] Great progress in understanding multiferroicity in orthorhombic system was achieved via far-infrared (IR) optical studies of electromagnons.[13,14,15,16,17,18] In contrast, far-IR optical excitations in hexagonal manganites are less understood. The recent studies have been mostly limited to YMnO$_3$ and HoMnO$_3$.[19,20] This paper presents spectra of magnons and crystal field (CF) excitations for a number of hexagonal manganites with ($RE$= Er, Tm, Yb, and Lu). Our experimental data will be discussed in comparison with the previously published far-IR results for one of the most studied hexagonal compound, HoMnO$_3$.[20]

Hexagonal manganites $RE$MnO$_3$ ($RE$= Ho, …Lu, and Sc, Y) exhibit ferroelectric (FE) order with a fairly large remnant polarization at high temperatures with the $T_C$ values in the range between 600 K and 1000 K.[21,22] The $RE$MnO$_3$ hexagonal structure consists of close-packed layers of MnO$_5$ bipyramids, which share corners in the $a$-$b$ planes. Along the $c$-axis, the layers of MnO$_5$ are well separated by the $RE^{3+}$ ions. A cooperative tilting of the bipyramidal sites below $T_C$ displaces the $RE^{3+}$ ions along the $c$-axis into two non-equivalent 2$a$ and 4$b$ sites. The oxygen ions are also displaced in the $a$-$b$ plane. Both displacements of $RE^{3+}$ ions and oxygen result in the FE polarization.[23,24] Detailed drawings of the hexagonal $RE$MnO$_3$ crystal structure [the polar space group $P6_3cm$ ($C_{6v}^3$) ] along with more detailed discussion of the ionic displacements in the FE phase can be found in Refs. 2, 23, 25.

The magnetic structure of hexagonal manganites $RE$MnO$_3$ have been studied in a number of publications.[25,26,27,28] However, the most intriguing part about the magnetic interaction between Mn$^{3+}$ and $RE^{3+}$ spins at low temperatures and high magnetic fields is still under debate.[5,29] The commonly accepted view on the magnetic structure and the corresponding magnetic phase transitions is the following. An antiferromagnetic (AFM) order of the Mn$^{3+}$ spins occurs at much lower temperatures compared to the FE transition. The AFM transition temperature $T_N$ for Mn$^{3+}$ spins is in the range between 70 K and 87 K for $RE$= Y, Ho, Er, Tm, Yb, and Lu depending almost linearly on $r_i$.[2] The neighboring spins of the close-packed Mn$^{3+}$ ions are AFM-coupled via the oxygen ions by superexchange interaction, which gives rise to frustration effects of an ideal 120° angle structure with the space group $P6'_3 c'm$. The Mn$^{3+}$ spins are ordered perpendicular to the $c$-axis: $\vec{S}_{Mn} \perp c$, while at low temperatures spins of $RE^{3+}$ ions ($RE$=Ho, Er, Tm, Yb) are oriented along the $c$ axis: $\vec{S}_{RE} \parallel c$. $RE^{3+}$ ion spins $S_{RE}$ can interact among themselves and with the Mn$^{3+}$ spins. These interactions result in a complex phase diagram in the temperature-magnetic field parameter space $T-H$.[29] Among all hexagonal $RE$MnO$_3$ compounds with $S_{RE} \neq 0$, HoMnO$_3$ is the most studied material. Its magnetic structure is particularly interesting since it shows two additional phase transitions below $T_N$. Mn$^{3+}$ spin reorientation occurs at $T_{SR} \approx 40$ K and AFM ordering of Ho$^{3+}$ spins takes place at $T_{RE} \approx 5$ K, as observed in neutron scattering[30,31] and second-harmonic generation optical experiments.[32,33,34,35] The spin reorientation is believed to be related to the $S_{Mn}$ rotation in the $a$-$b$ plane by 90°, changing the magnetic symmetry from $P6'_3 c'm$ to $P6'_3 cm'$. At much lower temperatures $T < T_{RE}$, another modification of the Mn spin structure occurs restoring the $P6_3 cm$ symmetry. Both low-temperature transitions at $T_{SR}$ and $T_{RE}$ are also accompanied by a complete or partial ordering of the Ho$^{3+}$ spins, which structure is not resolved yet. As mentioned in Ref. [5], two possibilities are discussed in literature for the spins of two non-equivalent Ho$^{3+}$ sites: (*i*) Ho spins on the 4$b$ site develop AFM order below $T_{SR}$ while Ho spins on the 2$a$ site remain PM, and (*ii*) all Ho spins develop AFM order below $T_{SR}$. In any case, there is an agreement that the Ho sublattice exhibits long-range AFM order along the $c$-axis below $T_{RE} \approx 5$ K . [5]. Magnetization of the $RE$ spins at low temperatures in other hexagonal manganites with $RE$= Er, Tm, Yb has been also studied in Refs. 36, 37, 38, where Dzyaloshinskii–Moriya (DM) interaction[39,40] has been proposed as one of the mechanisms for coupling between *RE* spins with the partial AFM order along the *c*-axis and Mn spins that are ordered in the *a-b* plane.

Both far-IR optical and neutron scattering experiments can provide valuable information about spectra of magnons and crystal field (CF) transitions in hexagonal manganites. Inelastic neutron scattering revealed three magnon branches in YMnO$_3$, one of them is acoustic and the other two are degenerate at the center of the Brillouin zone and have frequency near 40 cm$^{-1}$ at $\vec{q} = 0$.[7,41,42,43] The AFM resonance was also observed in YMnO$_3$ at the same frequency by means of far-IR spectroscopy.[44] Recent polarized inelastic neutron scattering studies revealed that the excitation seen at 1.5 K near 40 cm$^{-1}$ has a hybrid character of magnetic spin wave and a lattice vibration.[45] The authors proposed to explain this mode hybridization by DM interaction. In the recent optical transmission measurements in HoMnO$_3$, Talbaev *et al.*[20] demonstrated the magnon frequency renormalization and enhancement of the magnon splitting in the external magnetic field and related these effects to superexchange interaction between Ho$^{3+}$ and Mn$^{3+}$ spins. The effective spin Hamiltonian for the Ho ion ground state was also determined. Below, we focus on hexagonal manganites with *RE*= Er, Tm, Yb, and Lu and particularly on the relationship between their far-IR optical excitations, such as magnons and CF excitations, and the magnetic phase transitions at low temperatures and in high magnetic fields.

II. SAMPLES and EXPERIMENTAL TECHNIQUES

The high-temperature flux growth technique was utilized to produce bulk crystals of *RE*MnO$_3$. Single crystal platelets with the hexagonal *c*-axis perpendicular to the surface and a cross section area of about 4×4 mm$^2$ and the natural thickness of about 0.1 mm were used for transmission measurements. The opposite sides of the sample were not wedged, resulting in relatively strong thickness interference fringes in the measured optical spectra. Some crystals had terraces at the surface that resulted to an effective thickness variation across the light beam and irregularities in the thickness fringes in transmission spectra. Conventional polishing of our samples is detrimental to their optical properties and was not implemented in this paper. The transmission experiments were carried out at the National Synchrotron Light Source,

Brookhaven National Laboratory, at the U4IR beamline equipped with a Bruker IR spectrometer and a LHe-pumped (~1.6 K) bolometer. Far-IR transmission spectra were measured using linearly polarized synchrotron radiation with the resolution of 0.3 cm$^{-1}$ in the spectral range between 8 and 125 cm$^{-1}$. Polarization of the transmitted light was not analyzed. An external magnetic field of up to 9 T was applied in the Faraday configuration, so that the directions of the light propagation and the field coincided with the *c*-axis of the crystals. Correspondingly, the electric and magnetic fields of light were always in the hexagonal plane perpendicular to the *c*-axis. The raw data of transmitted intensity were normalized to transmission through an empty aperture with the size equal to that of the sample. In some figures for thin samples with strong thickness interference fringes we re-normalized the transmitted intensity for that measured at high temperature or high magnetic field.

III. SPECTRA of MAGNONS and CRYSTAL FIELD EXCITATIONS in ZERO MAGNETIC FIELD

Figure 1(a,b,c,d) shows transmission spectra measured at $T$ = 4.5 K in four *RE*MnO$_3$ samples with *RE*=Er, Tm, Yb, and Lu. The absorption minima, which correspond to the IR-active modes, such as magnons (AFM resonances) and CF transitions are marked with arrows. The absorption lines in Fig. 1(a,b,c) are quite weak and, correspondingly, the shape of the transmission spectra is strongly affected by the broad interference fringes with the period of ~12 cm$^{-1}$ for LuMnO$_3$ and YbMnO$_3$ and ~25 cm$^{-1}$ for TmMnO$_3$. In contrast, a strong absorption background due to the CF transitions in ErMnO$_3$ suppresses the interference fringes, which are visible only in the narrow spectral range around 30 cm$^{-1}$. In the following, we will present corrected transmission intensity data for LuMnO$_3$ and YbMnO$_3$, where the interference fringes will be removed for clarity. Note that using our far-IR ellipsometry setup, we confirmed that the lowest frequency $a-b$ plane polarized IR-active optical phonon in hexagonal *RE*MnO$_3$ is at ~160 cm$^{-1}$, which is well above the spectral range shown in Figure 1.

LuMnO$_3$: The transmission spectrum of LuMnO$_3$ shown in Fig. 1(a) has only one magnon peak positioned at 50.1 cm$^{-1}$. Since Lu$^{3+}$ $\left(^1S_0\right)$ has a complete $4f^{14}$ shell and zero spin, its compounds should not have any Lu-related crystal field electronic transitions in the far-IR spectral range. LuMnO$_3$ can be considered as a reference case for other *RE* hexagonal manganites where the optical properties in the far-IR spectral range are determined by Mn$^{3+}$ only. One can also expect that the IR optical properties of LuMnO$_3$ should be similar to that of the other hexagonal manganites without 4*f* electrons, such as YMnO$_3$ [44].

YbMnO$_3$: The transmission spectrum of YbMnO$_3$ shown in Fig. 1(b) has a magnon peak positioned at 53.1 cm$^{-1}$. Since Yb$^{3+}$ $\left(^2F_{7/2}\right)$ has incomplete $4f^{13}$ shell, a number of CF transitions can be expected. One CF peak related to the splitting of the ground level is at 11.5 cm$^{-1}$.

TmMnO$_3$: The transmission spectrum of TmMnO$_3$ shown in Fig. 1(c) has a magnon peak positioned at 49.8 cm$^{-1}$. Tm$^{3+}$ $\left(^3H_6\right)$ has incomplete $4f^{12}$ shell. The first CF transition is at 73.7 cm$^{-1}$. A spectral anomaly, which was detected at ~13 cm$^{-1}$ [not labeled in Fig. 1(c)], does not show any strong temperature or magnetic field dependences. We can attribute this feature at ~13 cm$^{-1}$ to the interference effects inside the sample, but additional studies of the samples with different thickness and larger in-plane cross section may be needed to confirm this assumption.

ErMnO$_3$: The transmission spectrum of ErMnO$_3$ shown in Fig. 1(d) has a magnon peak positioned at 47.5 cm$^{-1}$. Er$^{3+}$ $\left(^4I_{15/2}\right)$ has incomplete $4f^{11}$ shell. Since three *f* electrons are missing, the number of the CF transitions increases compared to that in other *RE*MnO$_3$ compounds. The transmission spectrum has several CF transitions: at 14 cm$^{-1}$ (CF1), and three closely spaced doublets at: 56.7 and 62.3 cm$^{-1}$ (CF2), 72 and 77.4 cm$^{-1}$ (CF3), 97 and 99 cm$^{-1}$ (CF4). In addition to that, we can identify two weaker spectral features at 22 and 40 cm$^{-1}$, which do not show any changes correlating with temperature or magnetic field. This allows us to attribute them to trivial thickness fringes.

To determine magnon and CF frequencies $\Omega_j$, adjusted oscillator strengths $A_j$, and broadening $\gamma_j$ parameters, the transmission spectra were fitted using a multi-oscillator dielectric model. The results of the fit are shown in Fig. 1 with solid black curves. An effective transmittivity function $T[\varepsilon(\omega) \cdot \mu(\omega)]$ was used for the fit. The corresponding electric and magnetic responses $\varepsilon(\omega)$ and $\mu(\omega)$ can be presented as a set of Lorentz oscillators

$$\varepsilon(\omega) = \varepsilon_\infty + \sum_{j=1}^{N} \frac{A_{j,e} \cdot \Omega_{j,0}^2}{\Omega_{j,0}^2 - \omega^2 - i\gamma_j \omega}$$

$$\mu(\omega) = \mu_\infty + \sum_{j=1}^{N} \frac{A_{j,m} \cdot \Omega_{j,0}^2}{\Omega_{j,0}^2 - \omega^2 - i\gamma_j \omega} \tag{1}$$

where $A_e$ is the oscillator strength for an electric dipole and $A_m$ is that for a magnetic dipole, and the off-resonance values of $\mu_\infty$ is assumed to be close to 1 [see Ref. 46 for more details about transmission spectra analysis]. As determined from the fit, the experimental values of $\varepsilon_\infty$ are close to 16.5, which was also confirmed with our ellipsometry measurements. The magnon parameters, calculated in the assumption that the sharp magnon peaks contribute only to $\mu(\omega)$ only are summarized in Table I. Note that the corresponding values of $A_m$ for $LuMnO_3$ and $YbMnO_3$ are smaller than that for $ErMnO_3$ and $TmMnO_3$. This difference can be due to the interaction between the magnons and the neighboring CF transitions and the difference in RE ion magnetization at low temperatures.

The temperature dependence for the magnon frequency $\Omega_M$ is shown in Fig. 2(a) for five $REMnO_3$ samples. The corresponding literature values of $T_N$ are shown with vertical arrows and are also listed in Table I. As expected, $\Omega_M(T)$ in $LuMnO_3$ demonstrates saturation for $T < 30$ K. When $T$ is approaching the AFM transition at $T_N \cong 87$ K, the magnon peak becomes weaker and its frequency decreases to ~45 cm$^{-1}$. Since we did not observe a complete softening of $\Omega_M$ even close to $T_N$, $\Omega_M(T)$ cannot be used to confirm the exact value of $T_N$. The experimental data for $\Omega_M(T)$ in $LuMnO_3$ were fitted using an empirical formula $\Omega_M(T) - \Omega_M(0) \sim T^\alpha$, where $\alpha = 3 \pm 0.5$. This temperature dependence of $\Omega_M(T)$ is similar to that for $YMnO_3$, where $\alpha = 3.5 \pm 0.5$.[44] The general trend for decreasing magnon frequency with temperature is also valid for the other four compounds with an incomplete $f$-shell, or $S_{RE} \neq 0$. Comparing the magnon frequency $\Omega_M(T = Const)$ for all 5 samples at a fixed temperature for $T \approx 25$ K, we observed that the magnon frequency decreases systematically with the increase of $r_i$ from Lu to Ho. At lower temperatures, however, the expected trend is violated and one can see a sudden increase of the magnon frequency upon cooling in RE compounds with an incomplete $f$ shell. For example, for $T < 20$ K, the $\Omega_M(T)$ curve for YbMnO$_3$ is above that for LuMnO$_3$ : $\Omega_M^{Yb} > \Omega_M^{Lu}$. We relate this effect to the increasing interaction between Mn$^{3+}$ and PM-ordered $RE^{3+}$ spins for $T < 25$, which will be discussed in the next *Sections*. For temperatures above ~ 20 K, the magnon frequency in ErMnO$_3$ fits well with the same function $\Omega_M^{Er}(T) - \Omega_M^{Er}(0) \sim T^\alpha$ as that for LuMnO$_3$. However, below $T_{SR}$ we can see a clear deviation from this classical behavior. One of the four measured $RE$MnO$_3$ samples, YbMnO$_3$, shows an unusual dependence of the magnon spectrum below $T < T_{RE}(Yb) = 3.3$ K, where a single magnon line splits into several distinct lines centered around 53 cm$^{-1}$. In the next section we will attribute this effect to the AFM ordering of Yb spins for $T < T_{RE}(Yb)$.[36]

The temperature dependence of the CF transitions in Er and Tm compounds is rather flat. As shown in Fig. 3(a,b) the CF peaks become weaker and broader with the temperature increase up to $T$=100 K due to the interaction of $f$ electrons with acoustic phonons and thermal depopulation of the ground state. One of the possible schematics of the CF transitions in ErMnO$_3$ that was derived from the corresponding transmission data in Fig. 3(b) is shown in Figure 4. Note however, that we did not take into account the possibility that the CF levels is not the same for Er ions in two sites of the lattice. Thus, CF2, CF3, and CF4 splitting could be due to this difference.

IV. SPECTRA of the AFM RESONANCES and CF EXCITATIONS in MAGNETIC FIELD

In external magnetic field $H$ directed along the $c$-axis, the doubly-degenerate magnon splits into two branches according to its effective $g$-factors. Figure 5(a) shows such dependence of the magnon frequency for LuMnO$_3$ *vs.* magnetic field: $\Omega_M^{\pm}(H) = \Omega_M(0) \pm \tfrac{1}{2} g \mu_B H$, where $\mu_B$ is the Bohr magneton ($\mu_B \approx 0.4669$ cm$^{-1}$/T). As expected for Lu$^{3+}$ compounds with $S_{RE} = 0$, the magnon $g$-factor is determined by Mn$^{3+} -$Mn$^{3+}$ interaction only. From the linear fit of the magnon doublet frequencies, we determined the experimental value of the $g$-factor to be $2.15 \pm 0.1$, which is close to the theoretical value of the Mn$^{3+}$ $g$-factor: $g_{Mn} = 2$. If we fit the high-magnetic field data points for the magnon doublet for $H > 2$ T, where the double splitting is much better defined, then the experimental value for the magnon $g$-factor becomes even closer to the theoretical expectation: $g = 2.05 \pm 0.05$. The temperature dependence of the magnon splitting in LuMnO$_3$ is negligible up to $T$=30 K. At higher temperatures the magnon line becomes broader, but still no substantial temperature dependence of the $g$-factor can be observed within the accuracy of our measurements.

In external magnetic field, the magnon lines in $RE$MnO$_3$ compounds with non-zero spin $S_{RE}$ behave differently compared to that for LuMnO$_3$. First, the splitting $\pm \frac{1}{2} g \mu_B H$ is enhanced corresponding to increases in the effective $g$-factor well above $g_{Mn} = 2$. The magnon $g$–factor values for all compounds, which have been determined in relatively weak magnetic fields ($H < 3$ T), are listed in Table I. One can see that $g$-factors vary non-systematically between 3.3 and 5.7 for different $RE$ hexagonal manganites. In contrast to the temperature-independent $g$-factor for LuMnO$_3$, the low-field values of the magnon $g$-factor in YbMnO$_3$, Tm MnO$_3$, and ErMnO$_3$ strongly depend on temperature as shown in Figure 6. For example, the low-temperature value in YbMnO$_3$ is $g = 5.7$, while at higher temperatures $T > 30$ K, the magnon $g$-factor decreases by a factor of approximately three, approaching the same value as we observed in LuMnO$_3$: $g(T) \to 2$. For $T > 40$ K the magnetic splitting vanishes and the broad magnon peak shows a rather weak quadratic increase of its frequency between 45 cm$^{-1}$ at zero field up to ~47 cm$^{-1}$ for $H = 2$ T. This behavior allows us to conclude that the interaction between AFM ordered Mn$^{3+}$ spins with $RE^{3+}$ spins quickly vanishes with the temperature increase.

Another interesting feature that can be observed in all compounds with a non-zero spin of $RE^{3+}$ ions is a sudden increase of the magnon doublet frequencies $\Delta$ that occurs at a certain value of a critical field $H_C \approx 4$ T. This increase is shown in Fig. 5(b) for TmMnO$_3$ and can be described as follows:

$$\Omega_M^\pm(H) = \Omega_M(0) \pm \tfrac{1}{2} g \mu_B H \qquad H < H_C \approx 4\text{ T}$$

$$\Omega_M^+(H) = \Omega_M(0) + \Delta + \tfrac{1}{2} g' \mu_B H \qquad H > H_C \approx 4\text{ T} \qquad (2)$$

$$\Omega_M^-(H) = \Omega_M(0) + \Delta - \tfrac{1}{2} g' \mu_B H$$

The high field values of $g'$ are listed in Table I for all samples. Typical values of $\Delta$ are about 2 cm$^{-1}$. Taking into account the experimentally determined value for the effective magnon *g*-factor (see Table I), $\Delta$ can be attributed to the appearance of an internal magnetic field $B_{INT} = 2\Delta/(g\mu_B)$, which is about 3.5 T in TmMnO$_3$. It is also important to note that the high-field values of the effective *g*-factors for the magnon peaks measured separately for $H > H_C$ are significantly lower than that determined at the low fields (see Table I). The $g'$ values for $H > H_C$ in $RE$MnO$_3$ ($RE$=Yb, Tm, Er) vary between 2.1 and 2.6 (much closer to $g_{Mn} = 2$), while the low-field values, which we discussed above, are significantly larger. In the following *Section* we will discuss this effect in more detail and will attribute it to magnetic-field induced saturation for $RE^{3+}$ magnetization for $H > H_C$.

In the previous section [see Fig. 3] we mentioned that in YbMnO$_3$, the magnon splits into several peaks at low temperatures. A closer view at this effect is presented in Fig. 7(a,b,c). Note that this effect was not observed in other measured $RE$MnO$_3$ samples probably due to lower values of $T_{RE}$. Temperature dependence of the transmission intensity in YbMnO$_3$ measured at zero magnetic field is shown in Fig. 7(a). At higher temperatures [left-hand side of Fig. 7(a)] one can see a single magnon at ~53 cm$^{-1}$, but for $T < T_{RE} = 3.3\text{ K}$ [right-hand side of Fig. 7(a)] this magnon splits into three weaker lines. The gaps between these three lines, which increase with the temperature decrease, are marked as $\Delta_{13}$ and $\Delta_{23}$. The maximum value of $\Delta_{13}$ is about 4 cm$^{-1}$, while the splitting of the upper magnon branch $\Delta_{23}$ reaches only about 2 cm$^{-1}$ at low temperatures. A much smaller splitting of the lower magnon branch at $T \approx 3$ K is difficult to analyze due to insufficient spectral resolution. Both gaps, $\Delta_{13}$ and $\Delta_{23}$, may be attributed to the AFM alignment of Yb$^{3+}$ spins and the corresponding changes in the internal magnetic fields $B_{13}$ and $B_{23}$ that affect Mn$^{3+}$ spins. To estimate the internal field values $B(T) = \Delta(T)/(g''\mu_B)$, we need to know the effective magnon *g*-factors, which are not necessarily the same as that measured in YbMnO$_3$ at $T > T_{RE}$. Indeed, our magnetic field measurements demonstrated that the effective magnon g-factor reduces from 5.8 to 2 for $T < T_{RE}$. Figure 7(b) shows the magnetic field dependence of the transmission intensity measured at $T = 1.45$ K, which corresponds to the right edge of the transmission intensity map in Figure 7(a). At low field, all three major magnon branches experience linear shifts with $H$ as follows: $\Delta_{13}(T,H) = g''\mu_B [B_{13}(T) + H]$ and $\Delta_{23}(T,H) = g''\mu_B [B_{23}(T) + H]$. The experimental values of $g'' = 2.0 \pm 0.1$ are the same for both gaps and coincide with $g_{Mn}$ within the accuracy of our measurements. The estimated values for $B_{13}$ and $B_{23}$ are plotted in Figure 7(c). Their maximum values are about 4.5 T and 2 T, respectively. It is interesting to mention that the external magnetic field $H \approx B_{23}$ and $H \approx B_{13}$ results in two drastic changes in the magnon g-factor value. Figure 7(b) shows a sudden change in both $\Delta_{13}(H)$ and $\Delta_{13}(H)$ dependencies. Between $H = 2.3$ T and $H = 5$ T, the magnon g-factor is $g'' = 4$, while for $H > 5$ T the g-factor value reduces back to $g'' \approx 2.1$.

The CF peaks in Yb, Tm, and Er compounds show strong magnetic field dependencies as shown in the corresponding 2D transmission intensity maps in Figure 8(a,b,c). The common behavior for all three compounds reveals itself in a linear increase of the lowest CF frequencies in a magnetic field and in the change in the slopes for the magnetic field above $H_C$. In Fig. 8(a) one can see a 2D transmission intensity map for YbMnO$_3$. For $H < H_C \approx 4$ T, a linear increase of the CF transition can be described as $\Omega_{CF}(H) = \Omega_{CF}(0) + g_{Yb}\mu_B H$ with $g_{Yb} = 2.8$. Similarly to that for magnons, the CF frequency for $H > H_C$ has a jump and a reduced slope, which can be described by $\Omega_{CF}(H) = \Omega_{CF}(0) + \delta + g'_{Yb}\mu_B H$ with $g'_{Yb} \approx 2$ and $\delta \approx 5$ cm$^{-1}$. The single CF line in TmMnO$_3$ seems to be unchanged for $H < H_C \approx 3.5$ T and it shifts up linearly with field for $H > H_C$. The most obvious magnetic-field-induced changes in the spectra of the CF transitions have been observed in ErMnO$_3$. The 2D transmission map is shown in Figure 8(c). One can see a linear variation of the CF lines at low magnetic field according to their effective g-factors. The CF2 and CF 3 doublets split in magnetic field $H < H_C$ with g-factors of about $g_{Er(CF2,CF3)} = \pm 4.5$. Above a certain value of the critical field $H_C$, which turns out to be temperature-dependent, most of the CF lines experience either a frequency jump or a sudden strong splitting, like the CF4 doublet at 98 cm$^{-1}$. The higher frequency doublet C4 and the lower frequency level CF1 do not show any action at weak magnetic fields ($g_{Er(CF1,CF4)} \approx 0$). For $H > H_C$, the high frequency doublet CF4 splits with the same g-factor $g'_{Er(CF4)} = \pm 4.5$ and CF1 increases its frequency with $g'_{Er(CF1)} = 1.5$. Figure 9 shows an increase of $H_C$ that varies between ~1.7 T for $T = 3.5$ K and ~8 T at $T = 35$ K. In the next *Section* we will discuss the magnetic field- and temperature-induced changes in the frequencies of magnons and CF transitions and their *g*-factors for several *RE*MnO$_3$ compounds with and without *RE* spins.

V. DISCUSSION

In the temperature range $T \geq 25$ K, the magnon frequency in all measured samples appears to follow the empirical trend $\Omega_M(T) - \Omega_M(0) \sim T^\alpha$, which is determined by the interactions between Mn spins only. The corresponding values of the magnon frequencies $\Omega_M(T = 25 \text{ K})$ are summarized in Table I. The temperature $T = 25$ K is chosen in the following analysis for its optimum value: the Mn-*RE* interaction is already negligibly weak and the temperature is not high enough to affect the Mn-Mn spin interaction. In this narrow temperature range, the changes of the magnon frequency with $r_i$ may be related to the corresponding variation of the primary antiferromagnetic exchange constant $J$ and anisotropy $D$ for Mn spins. The corresponding Hamiltonian can be written as follows [43]

$$\hat{H} = J \sum_{i,j} \vec{S}_i \cdot \vec{S}_j + D \sum_i \left(S_i^z\right)^2, \tag{3}$$

where the sum is taken over nearest-neighbor inplane spin pairs. The magnon frequency can be expressed in terms of $J$ and $D$ constants: $\Omega_M = 3 S_{Mn} \sqrt{J \cdot D}$, where $S_{Mn} = 2$ is the spin of Mn.[20] The empirical values for the $J \cdot D$ product are shown in Table I demonstrating a systematic decrease of $J \cdot D$ with increase of $r_i$. As we will see in the following, the measurement of the magnon frequency at the zone center cannot always separate the $J$ and $D$ constants, even if the empirical magnetic-field dependence $\Omega_M^\pm(H)$ is considered.

The Hamiltonian for Mn spin interaction in external magnetic field $H$ directed along $c$-axis includes another term:

$$\hat{H} = J\sum_{i,j}\vec{S}_i \cdot \vec{S}_j + D\sum_i \left(S_i^z\right)^2 - \mu H \sum_i S_i^z. \tag{4}$$

Correspondingly, in the approximation of weak $H$, such as $H < S_{Mn}J \approx 30$ T, the quasi-linear splitting of the magnon branches $\Omega_M^\pm(H)$ can be written as follows [20,44,47]

$$\Delta\Omega_M^\pm(H) = \frac{g_{Mn}J}{J + 2D/9}\mu_B H = g\mu_B H. \tag{5}$$

For positive $J$ and $D$ constants, this model allows only solutions with the effective magnon $g$-factors of $g \leq g_{Mn} = 2$. When the $D/J$ ratio is less than 0.2, the calculated $g$-factor asymptotically approaches $g_{Mn} = 2$ and the fit for the magnon frequency becomes insensitive to the individual values of $J$ and $D$. Consequently, within the accuracy of our measurements, which show the empirical values of the magnon $g$-factor to be close to 2 for $T \geq 25$ K [see Fig. 6], we can only determine the $J \cdot D$ product, which is listed in Table I. Neutron scattering measurements of the magnon dispersion may help to determine $J$ and $D$ constants, as it has been done for HoMnO$_3$ in Ref. 43 (see Table I for the experimental values of $J$ and $D$).

The observed magnon frequency deviation from $\Omega_M(T) - \Omega_M(0) \sim T^\alpha$ with the temperature decrease below $T < 25$ K [Fig. 2(a)] is a strong indication of the Mn-RE spin interaction. The simplest way to describe the effect of $RE^{3+}$ ions is to consider an exchange interaction between Mn spins and the temperature-average spins of $RE^{3+}$. Magnetization of $RE$ ions can be estimated as $M_{RE} = g_{RE}J_{RE}\mu_B B_J(x)$, where the argument of the Brillouin function $B_J(x)$ $x = g_{RE}\mu_B J_{RE} B_{INT}/k_B T$ depends on the internal field $B_{INT}$. Figure 2(b) shows calculations for a relative change of magnetization $M_{Er}(T)/M_{Er}(0)$ for $Er^{3+}$ using parameters $J_{RE} = 15/2$, $g_{Er} = 1.2$. The value of internal field and $B_{INT} = 2$ T is chosen to fit the experimental data for $Er^{3+}$ magnetization from Ref. 38. The increase of the empirical values of the magnon frequency above the level shown with the red solid line $\Omega_M(T) - \Omega_M(0) \sim T^\alpha$ in Figure 2(a) for ErMnO$_3$ correlates well with the calculated dependence for $M_{Er}(T)/M_{Er}(0)$ in Figure 2(b). The effective increase of the magnon $g$-factor for the temperature decrease in the range $T < 25$ K (Fig. 6) can be also related to the PM contribution of $RE^{3+}$ ions. If we assume a ferromagnetic coupling between Mn and RE spins, then the magnetization component $M_{RE}^z$ along the external magnetic field ($H \parallel c$) is equivalent to an additional magnetic field $B_{RE} \sim M_{RE}^z(H,T) \sim H$ that influences the Mn spins and results in an increase of the effective magnon $g$-factor above 2: $(g - g_{Mn}) \sim M_{RE}^z(H,T)$ [see Fig. 6]. In the weak magnetic fields and $T < 25$ K, $M_{RE}^z(H,T) \sim H/T$, while above 25 K, $M_{RE}^z(H,T)$ approaches zero.

In all hexagonal compounds, we found that above a critical external field $H_C$, which is between 2.5 T and 4 T for different compounds at $T \approx 5$ K, the effective $g$-factor decreases suddenly to approximately 2. This effect can be qualitatively understood if we take into consideration two $RE$ sublattices: below $H_C$ two sublattices of $RE$ are AFM-coupled to each other, but above $H_C$ we have a complete reorientation of the $RE$ spins along the magnetic field. In this case the AFM interaction between $RE$ spins changes to FM and the paramagnetic $M_{RE}^z(H)$ reaches saturation and cannot contribute to the effective magnon $g$-factor, which becomes $g \approx g_{Mn}$.

Figure 9 shows the critical external field $H_C$ as a function of temperature for ErMnO$_3$. At low temperatures, $H_C$ increases linearly with $T$: $3\mu_B H_C \approx k_B T$. This behavior corresponds to the formal requirement to keep the PM magnetization constant $M_{RE}^z(H,T) \sim H/T \approx Const$. At low external fields, the crystal field that includes the Mn-$RE$ interaction dominates over the external perturbation. Two CF levels, CF1 and CF4 have zero $g$-factors along $c$-axis, while $g$-factors of CF2 and CF3 doublets are $g_{Er(CF2,CF3)} = \pm 4.5$. At high external fields, Er$^{3+}$ ions re-orient their magnetization along the $c$-axis that is reflected in the sudden change of $g_{Er(CF4)}$ from zero to $\pm 4.5$.

PM to AFM transition can explain the magnon behavior in YbMnO$_3$ for $T < 3.3$ K shown in Figure 7. The PM contribution to the effective magnon $g$-factor $(g - g_{Mn}) \approx 3.8 \pm 0.6$ suddenly decreases due to the transition between the PM and AFM ordering of Yb spins, $M_{Yb}^z$ reaches saturation for $T < 3.3$ K, and the magnon $g$-factor becomes $g \approx g_{Mn}$. In the high field $H > 5$ T the complete saturation of the $RE$ sublattice is achieved along the $c$-axis and $g(H > 5 \text{ T}) \approx g_{Mn}$.

To find the exact magnetic structure for $\{T < 3.3\ \text{K},\ 2\ \text{T} < H < 5\ \text{T}\}$ and to explain the effective magnon $g$-factor equal to 4 in this range, one needs to perform precise theoretical calculations for the interaction between the Mn and Yb spin sublattices.

VI. CONCLUSIONS and ASKNOWLEGEMENTS

We presented experimental results for the magnon and CF transmission spectra in several single crystal $RE$MnO$_3$ hexagonal manganites. The observed magnon frequencies increase systematically with the famous decrease of the $RE$ ionic radius $r_i$ between Ho and Lu. The measured magnon frequencies allow to determine the product of $J \cdot D$, but do not permit their separation due to the condition $J \gg D$ and high internal exchange fields between Mn spins. At low temperatures below ~30 K the Mn-$RE$ interaction changes both the magnon $g$-factors and the magnon frequencies. Both trends are explained qualitatively by the changes of $RE$ magnetization with temperature and magnetic field, where: $M_{RE}^z(H,T) \sim H/T$. In YbMnO$_3$ we observed the AFM ordering of Yb spins below $T_{RE} = 3.3$ K. In other compounds (TmMnO$_3$ and ErMnO$_3$) we did not see this effect probably due to even lower ordering temperatures for $RE$ spins. The parameters of the magnon and CF spectra can be useful for precise theoretical models for $RE$-Mn interaction in hexagonal manganites and for better understanding of the interplay between the AFM and FE effects in these compounds.

The authors are thankful to G. L. Carr for help at U4IR beamline, to Weida Wu for useful discussions, and to D. Talbayev for useful discussions and for providing the unpublished data for the temperature dependence of the magnon frequency in HoMnO$_3$. Work at NJIT and Rutgers was supported by DOE DE-FG02-07ER46382. Use of the National Synchrotron Light Source, Brookhaven National Laboratory, was supported by the U.S. Department of Energy, Office of Science, Office of Basic Energy Sciences, under Contract No. DE-AC02-98CH10886.

Table I. Magnon frequency, oscillator strength in units of $\mu_\infty$, and broadening: $\Omega_M$, $A_m$, and $\gamma$. The low field and high field values of the effective magnon g-factors: $g$ and $g'$ measured at T=4.5 K for $H \parallel c$. The magnon frequency $\Omega_M(T = 25\text{ K})$ that has been used to determine the $J \cdot D$ product.

| | $T_N$ [c] (K) | $\Omega_M$ (cm$^{-1}$) | $A_m$ $\times 10^{-3}$ | $\gamma$ (cm$^{-1}$) | $g$ ($H < H_C$) | $g'$ ($H > H_C$) | $\Omega_M$ $T = 25$ K (cm$^{-1}$) | $J \cdot D$ $T = 25$ K (cm$^{-2}$) |
|---|---|---|---|---|---|---|---|---|
| LuMnO$_3$ | 88 | 50.2 | 1.0 | 1.1 | 2.15±0.1 | 2.05±0.05 | 49.8 | 68.9 |
| YbMnO$_3$ | 84 | 53.0 | 1.0 | 0.6 | 5.8±0.6 | 2.65±0.15 | 49.6 | 68.3 |
| TmMnO$_3$ | 82 | 49.9 | 2.5 | 1.0 | 3.3±0.1 | 2.1±0.05 | 48.4 | 65.1 |
| ErMnO$_3$ | 78 | 47.0 | 4.0 | 1.1 | 4.4±0.3 | 2.5±0.2 | 45.6 | 57.8 |
| HoMnO$_3$ | 75 | 43.4 [a] | | | ~4 [a] | ~2.4 [a] | 41.7 [a] | 44.4 [d] $J = 19.7$ cm$^{-1}$ $D = 2.3$ cm$^{-1}$ |
| YMnO$_3$ [b] | 76 | 43 | | | 1.9±0.1 | | | |

[a] IR transmission data from Ref. 20
[b] IR transmission data from Ref. 44
[c] Magnetization data from Ref. 2
[d] Neutron data from Ref. 43

REFERENCES

[1] J. A. Alonso, M. J. Martinez-Lope, M. T. Casais, and M. T. Fernandez-Diaz, Inorg. Chem. **39**, 917 (2000).

[2] J.-S. Zhou, J. B. Goodenough, J. M. Gallardo-Amores, E. Mor´an, M. A. Alario-Franco, R. Caudillo, Phys. Rev. B **74**, 014422 (2006).

[3] Z. J. Huang, Y. Cao, Y. Y. Xue, and C. W. Chu, Phys. Rev. B **56**, 2623 (1997).

[4] J. C. Lemyre, M. Poirier, L. Pinsard-Gaudart, and A. Revcolevschi, Phys. Rev. B **79**, 094423 (2009).

[5] B. G. Ueland, J. W. Lynn, M. Laver, Y. J. Choi, and S.-W. Cheong, Phys. Rev. Lett. **104**, 147204 (2010).

[6] T. Goto, T. Kimura, G. Lawes, A. P. Ramirez, and Y. Tokura, Phys. Rev. Lett. **92**, 257201 (2004).

[7] S. Petit, F. Moussa, M. Hennion, S. Pailhès, L. Pinsard-Gaudart, and A. Ivanov, Phys. Rev. Lett. **99**, 266604 (2007).

[8] H. Katsura, N. Nagaosa, A. V. Balatsky, Phys. Rev. Lett. **95** 057205 (2005).

[9] I. A. Sergienko and E. Datotto, Phys. Rev. B **73** 094434 (2006).

[10] M. Mostovoy, Phys. Rev. Lett. **96** 067601 (2006).

[11] S.-W. Cheong and M. Mostovoy, Nature Materials **6**, 13 (2007).

[12] Y. Tokura, , J. Mag. Mag. Mat. **310**, 1145 (2007).

[13] A. Pimenov, A.A. Mukhin, V.Yu. Ivanov, V.D. Travkin, A.M. Balbashov, and A. Loidl, Nature Physics **2**, 97 (2006).

[14] A. Pimenov, A. M. Shuvaev, A. A. Mukhin, A. Loidl, J. Phys.: Condens.Matter **20** (2008) 434209.

[15] D. Senff, N. Aliouane, D. N. Argyriou, A. Hiess, L. P. Regnault, P. Link, K. Hradil, Y. Sidis, M. Braden, J. Phys.: Condens. Matter **20**, 434212 (2008).

[16] N. Kida, Y. Takahashi, J. S. Lee, R. Shimano, Y. Yamasaki, Y. Kaneko, S. Miyahara, N. Furukawa, T. Arima, Y. Tokura, J. Opt. Soc. Am. **B26,** A35 (2009).

[17] A. Pimenov, A. Shuvaev, A. Loidl, F. Schrettle, A. A. Mukhin, V. D. Travkin, V. Yu. Ivanov, and A. M. Balbashov, Phys. Rev. Let. **102**, 107203 (2009).

[18] A. M. Shuvaev, V. D. Travkin, V. Yu. Ivanov, A.A. Mukhin, and A. Pimenov, , Phys. Rev. Lett. **104**, 097202 (2010).

[19] C. Kadlec, V. Goian, K. Z. Rushchanskii, P. Kužel, M. Ležaić, K. Kohn, R. V. Pisarev, and S. Kamba, Phys. Rev. B **84**, 174120 (2011).

[20] D. Talbayev, A. D. LaForge, S. A. Trugman, N. Hur, A. J. Taylor, R. D. Averitt, D. N. Basov, Phys. Rev. Lett. **101**, 247601 (2008).

[21] S.C. Abrahams, Acta Crystallogr., Sect. B: Struct. Sci. **57**, 485 (2001).

[22] V. Goiana, S. Kamba, C. Kadleca, D. Nuzhnyya, P. Kuzela, J. Agostinho Moreirab, A. Almeidab and P.B. Tavares, Phase Transitions **83**, 931 (2010).

[23] M. N. Iliev, H.-G. Lee, V. N. Popov, M. V. Abrashev, A. Hamed, R. L. Meng, and C. W. Chu, Phys. Rev. B **56**, 2488 (1997).

[24] B. B. Van Aken, T. T. M. Palstra, A. Filippetti, and N. A. Spaldin, Nat. Mater. **3**, 164 (2004).

[25] I. Munawar and S. H. Curnoe, J. Phys.: Condens. Matter **18**, 9575 (2006).

[26] E.F. Bertaut and M. Mercier, Phys. Lett. **5**, 27 (1963).

[27] P. A. Sharma, J. S. Ahn, N. Hur, S. Park, Sung Baek Kim, Seongsu Lee, J.-G. Park, S. Guha, and S-W. Cheong, Phys. Rev. Lett. **93**, 177202 (2004).

[28] S. G. Condran and M. L. Plumer, J. Phys.: Condens. Matter **22** 162201 (2010).

[29] B. Lorenz, F. Yen, M. M. Gospodinov, and C. W. Chu, Phys. Rev. B **71**, 014438 (2005).

[30] A. Munoz, J. A. Alonso, M. J. Martı́nez-Lope, M. T.Casaı́s, J. L. Martı́nez, and M. T. Ferna´ndez-Dı́az, Chem. Mater. **13**, 1497 (2001).

[31] T. Lonkai, D. Hohlwein, J. Ihringer, and W. Prandl, Appl. Phys. A **74**, S843 (2002).

[32] M. Fiebig, D. Fro¨hlich, K. Kohn, S. Leute, T. Lottermoser, V.V. Pavlov, and R.V. Pisarev, Phys. Rev. Lett. **84**, 5620 (2000).

[33] M. Fiebig, D. Fro¨hlich, T. Lottermoser, and K. Kohn, Appl. Phys. Lett. **77**, 4401 (2000).

[34] M. Fiebig, C. Degenhardt, and R.V. Pisarev, J. Appl. Phys. **91**, 8867 (2002).

[35] M. Fiebig, D. Fro¨hlich, T. Lottermoser, and M. Maat, Phys. Rev. B **66**, 144102 (2002).

[36] X. Fabrèges, I. Mirebeau1, P. Bonville, S. Petit, G. Lebras-Jasmin, A. Forget, G. André, and S. Pailhès, Phys. Rev. B **78**, 214422 (2008).

[37] H. A. Salama, and G. A. Stewart. J. of Phys.: Condens. Mat. 21 (38), 386001 (2009).

[38] Yanan Geng, Nara Lee, Y. J. Choi, S-W. Cheong, Weida Wu, arXiv:1201.0694v1 [cond-mat.str-el].

[39] I. E. Dzyaloshinskii, Soviet Phys. JETP **10**, 628 (1960).

[40] T. Moriya, Physical Review **120** (1), 91 (1960).

[41] T. Chatterji, S. Ghosh, A. Singh, L.P. Regnault, and M. Rheinstadter, Phys. Rev. B, **76,** 144406 (2007).

[42] T. J. Sato, S.-H. Lee, T. Katsufuji, M. Masaki, S. Park, J.R.D. Copley, and H. Takagi, Phys. Rev. B **68,** 014432 (2003).

[43] O. P. Vajk, M. Kenzelmann, J.W. Lynn, S. B. Kim, and S.-W. Cheong, Phys. Rev. Lett. **94**, 087601 (2005).

[44] T Penney, P. Berger, and K. Kritiyakirana, J. Appl. Phys. **40**, 1234 (1969).

[45] S. Pailhes, X. Fabreges, L.P. Regnault, L. Pinsard-Godart, L. Mirebau, F. Moussa, M. Hennion, and S. Petit, Phys. Rev. B **79**, 134409 (2009).

[46] P. D. Rogers, Y. J. Choi, E. C. Standard, T. D. Kang, K. H. Ahn, A. Dudroka, P. Marsik, C. Bernhard, S. Park, S-W. Cheong, M. Kotelyanskii, and A. A. Sirenko, Phys. Rev. B **83**, 174407 (2011).

[47] W. Palme, F. Mertens, O. Born, B. Luthi, and U. Schotte, Solid State Commun. **76**, 873 (1990).

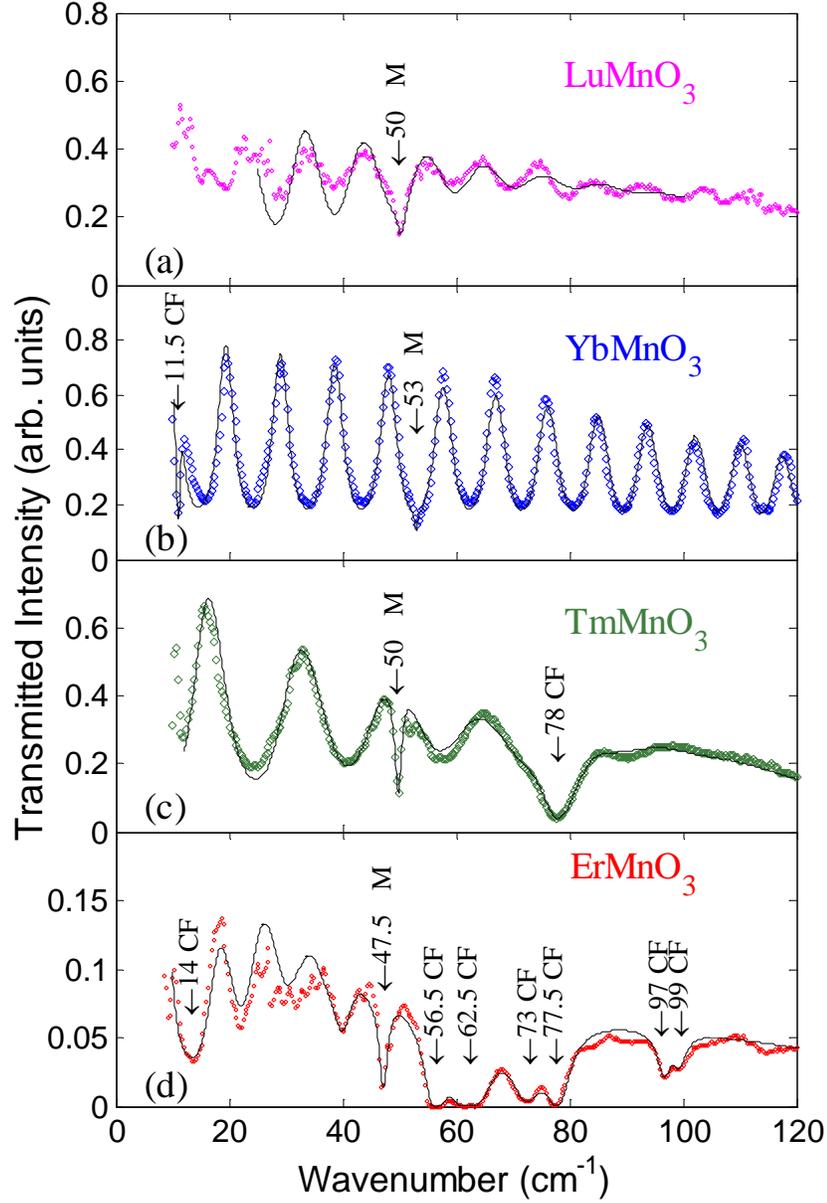

FIG. 1(color online)   Transmission spectra of *RE*MnO$_3$ samples measured for the light propagation along the *c*-axis at $T \approx 4.5$ K and at zero magnetic field. The magnon absorption lines are marked with M. The crystal field transitions are marked with CF. Black solid curves represent the fit results using the multi-oscillator dielectric model [see Eq.(1)]. The magnon and CF frequencies are rounded up to 0.5 cm$^{-1}$.

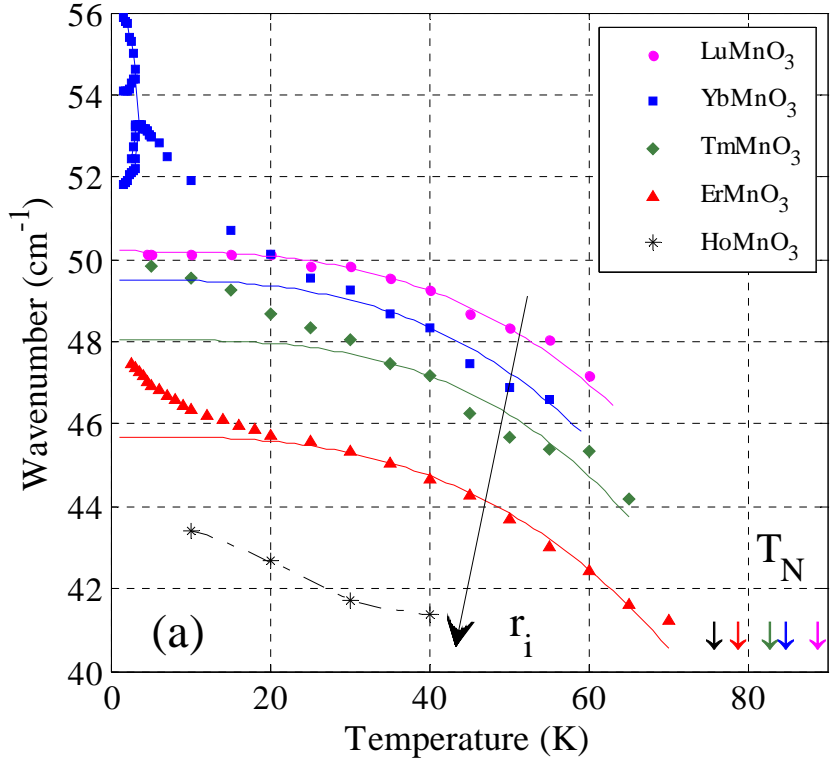

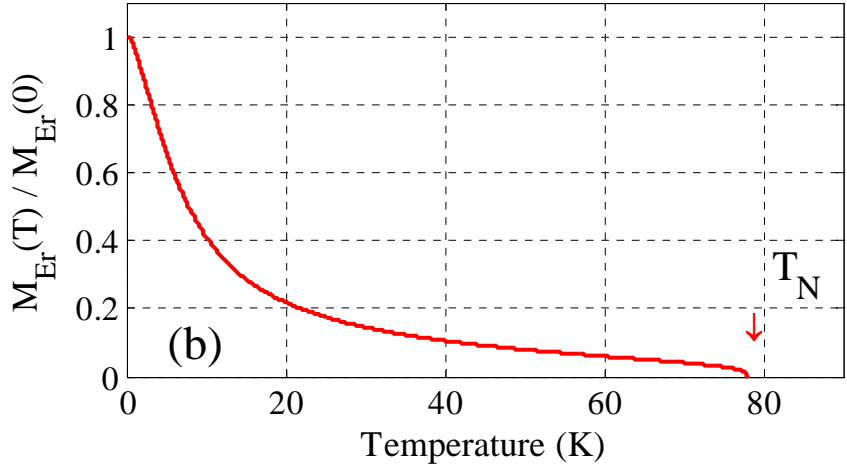

FIG. 2(color online) (a) Temperature dependence of the magnon frequency for five $RE$MnO$_3$ samples at zero magnetic field. Black stars correspond to HoMnO$_3$ (courtesy of D. Talbayev). At low temperatures, a single magnon in YbMnO$_3$ splits into several distinct absorption lines due to ordering of Yb spins at $T_{Yb} = 3.3\,\text{K}$. Solid curves for Lu, Yb, Tm, and Er samples show the fit results using $\Omega_M(0) - \Omega_M(T) \sim T^3$. Dashed curve for HoMnO$_3$ guide the eye. The AFM transition temperatures $T_N$ for Mn$^{3+}$ are shown with arrows. The arrow labeled $r_i$ indicates the increase of the $RE$ ionic radius and the corresponding decrease of both, magnon frequency and $T_N$. (b) Calculation for the relative change of paramagnetic magnetization of Er$^{3+}$.

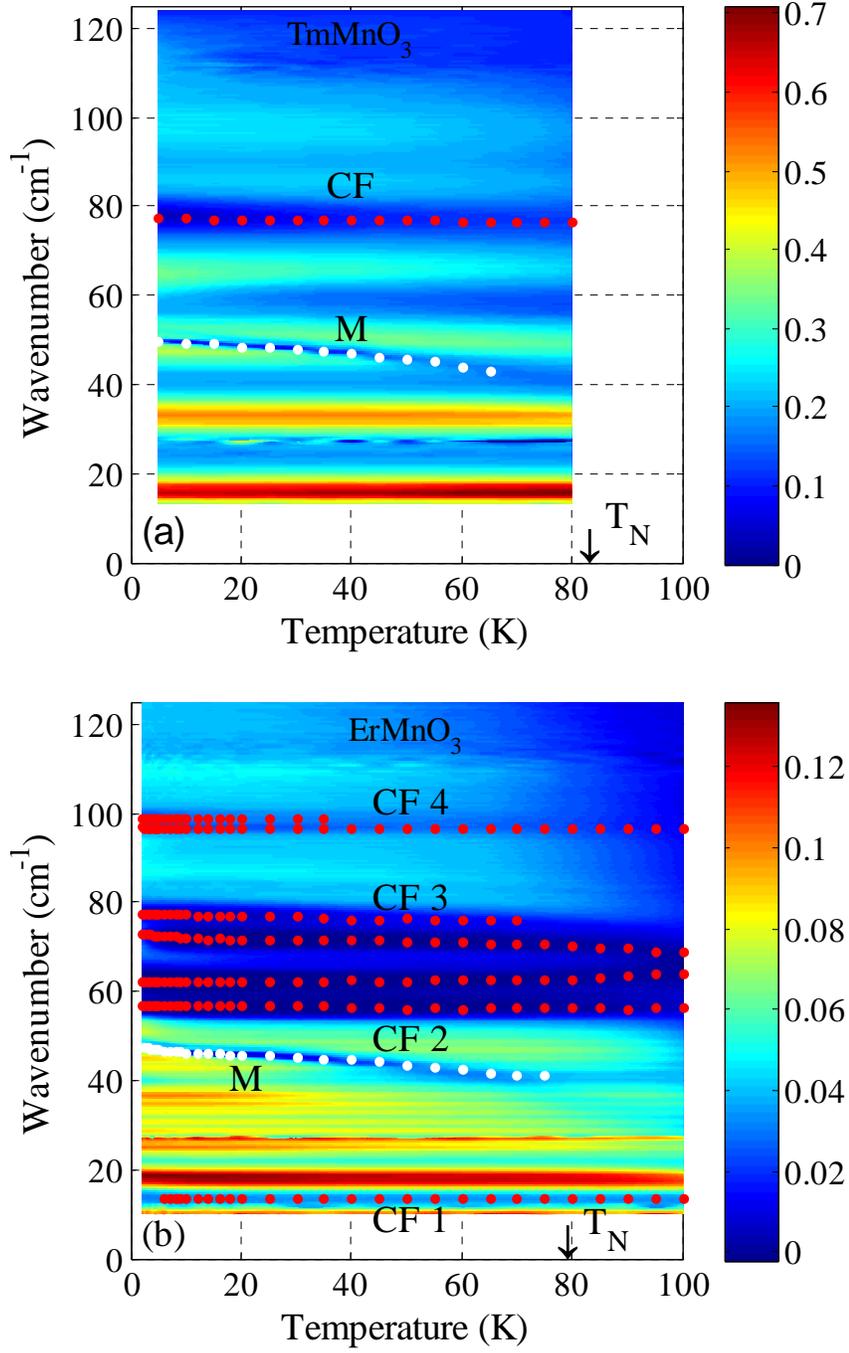

FIG. 3 (color online) Transmission maps *vs.* temperature and light frequency for TmMnO$_3$ (a) and ErMnO$_3$ (b). The blue (dark) color corresponds to stronger absorption and red (light) color indicates high transmission. The transmission intensity scale is shown with the vertical bars. Black arrows indicate the AFM transition temperatures $T_N$. Frequencies of the CF transitions are shown with red dots and the magnon frequencies are shown with white dots. Noise in the map (a) at ~ 27 cm$^{-1}$ is an artifact of the experimental setup.

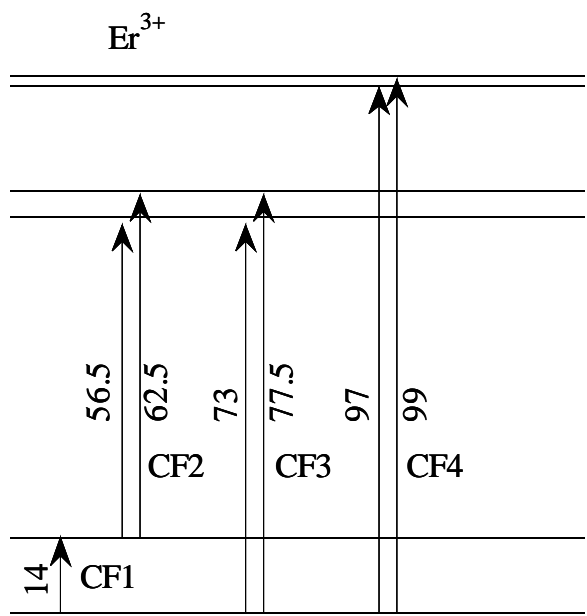

FIG. 4 Reconstruction of the CF transitions in $Er^{3+}$ in $ErMnO_3$, which is based on the corresponding transmission spectra in Fig. 1(d) and Fig. 3(b). The corresponding transition frequencies in $cm^{-1}$ are shown next to the vertical arrows rounded up to 0.5 $cm^{-1}$.

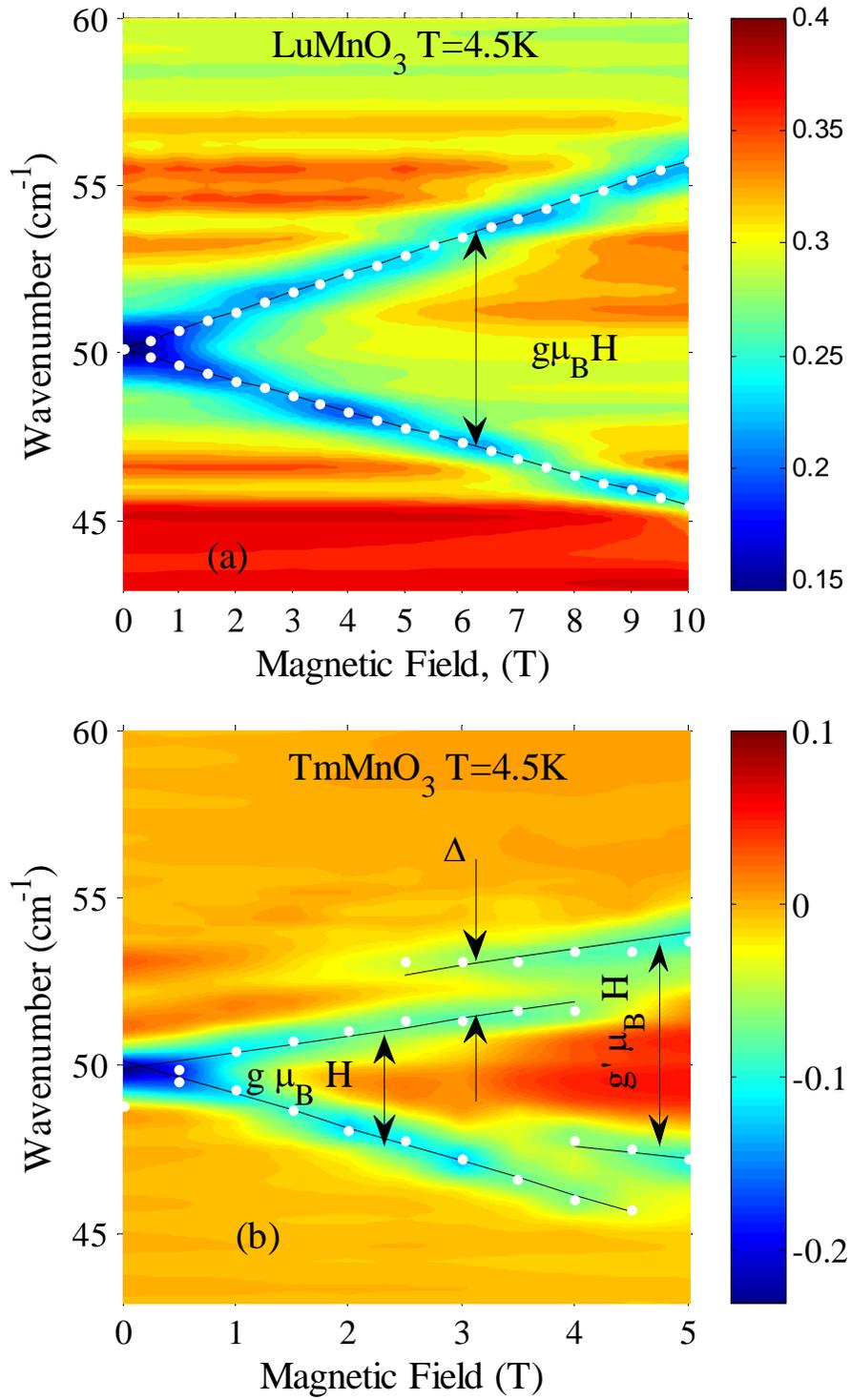

FIG. 5 (color online) Normalized transmission map *vs.* magnetic field and light frequency for $LuMnO_3$ (a) and $TmMnO_3$ (b). The blue (dark) color corresponds to stronger absorption and red (light) color indicates high transmission. The normalized transmission intensity scale is shown with the vertical bars. The white dots represent the fit results for the magnon doublet splitting in external magnetic field [see Eq.(2)].

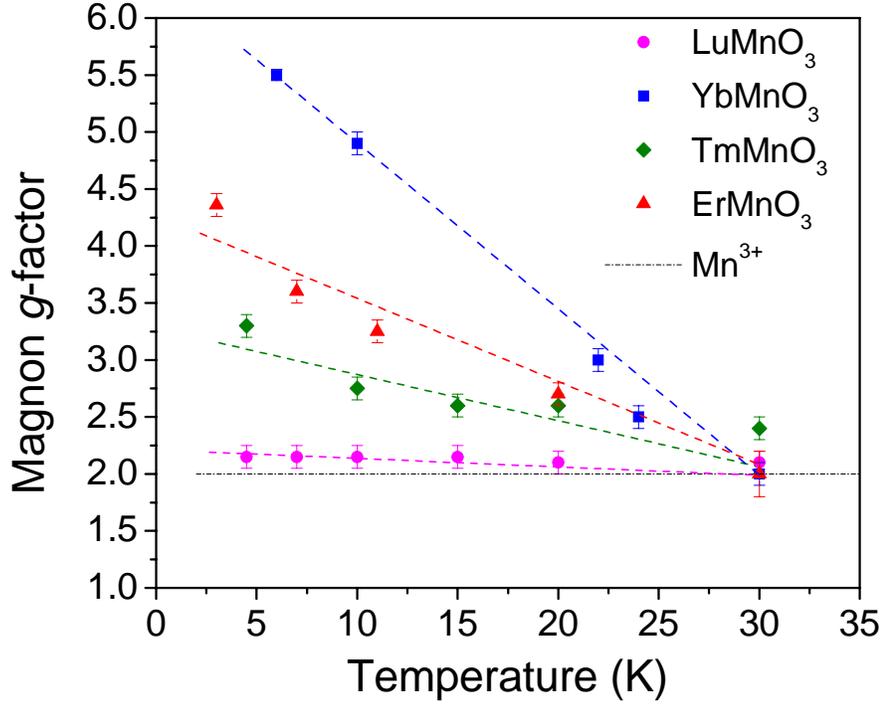

FIG. 6. Temperature dependence of the magnon $g$-factor for $RE$MnO$_3$ ($RE$=Lu, Yb, Tm, and Er). Dashed lines guide the eye. The horizontal line corresponds to the theoretical expectation for $g$-factor of Mn$^{3+}$: $g_{Mn} = 2$. At higher temperatures above $T \approx 30$ K, the magnon $g$-factors approach $g_{Mn} = 2$ in all $RE$MnO$_3$.

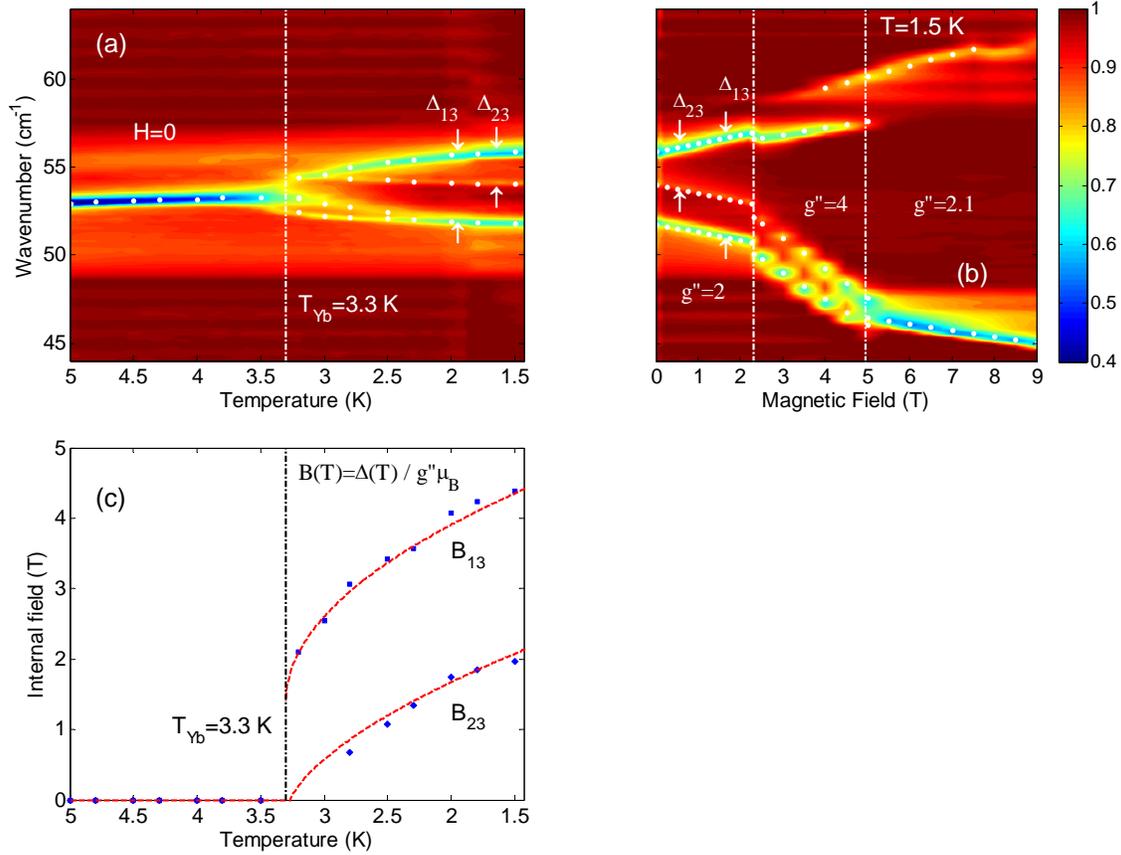

FIG. 7. (a) Temperature dependence of the transmission intensity in YbMnO$_3$ measured at zero magnetic field. The magnon lines are marked with white dots. Below $T_{Yb} = 3.3$ K the major splitting of the magnon is marked with $\Delta_{13}$. Two additional weak absorption lines can be seen with a smaller splitting $\Delta_{23}$ between the top two modes. (b) Magnetic field dependence of the transmission intensity measured at $T = 1.45$ K, which corresponds to the right edge of the map in (a). The vertical frequency scale and the intensity scale are the same for the maps in (a) and (b). Below $H = 2.3$ T, the major magnon doublet splits in external field as $\Delta_{13}(T,H) = g''\mu_B [B(T) + H]$ with $g'' = 2$. Between $H = 2.3$ T and $H = 5$ T, the magnon $g$-factor is $g'' = 4$. $g'' = 2.1$ for $H > 5$ T. (c) Temperature dependence of the two internal fields $B_{13}$ and $B_{23}$, which correspond to $\Delta_{13}$ and $\Delta_{23}$ splittings in (a) calculated as $B(T) = \Delta(T)/(g''\mu_B)$ with $g'' = 2$. Note that the reversed scale for temperature in (c) matches that in (a).

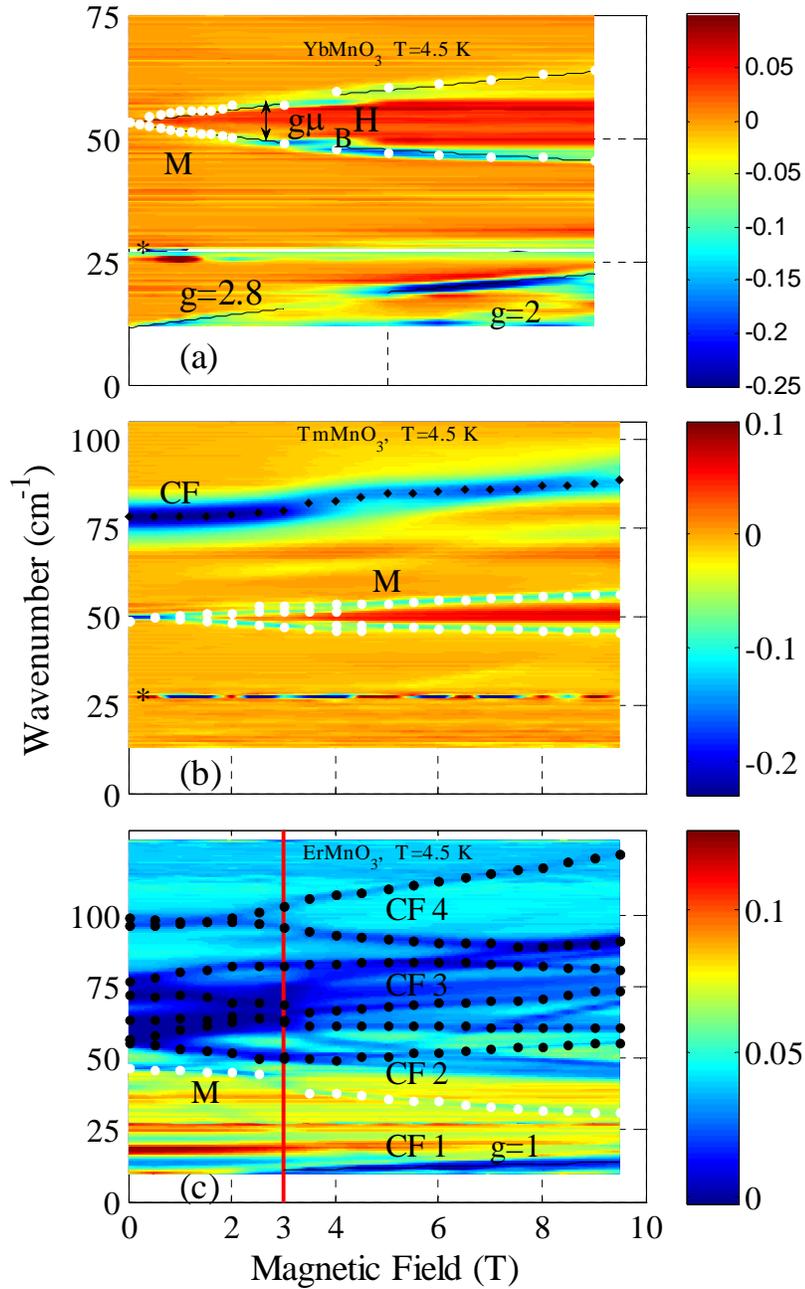

FIG. 8 (color online) Normalized transmission maps *vs.* magnetic field and light frequency for YbMnO$_3$ in (a), TmMnO$_3$ in (b), and ErMnO$_3$ in(c) measured at $T = 4.5$ K. The blue (dark) color corresponds to stronger absorption and red (light) color indicates high transmission. The transmission intensity scale is shown with the vertical bars. The black dots represent the fit results for the CF transitions. The magnons M are shown with white circles. Note the changes of the magnon g-factor for YbMnO$_3$ at the critical field $H_C \approx 4$ T in (a). The critical field for ErMnO$_3$, $H_C = 2.8$ T, is marked with the vertical red line. Noise in (a) and (b) maps at ~ 27 cm$^{-1}$, which is marked with (*) is an artifact of the experimental setup.

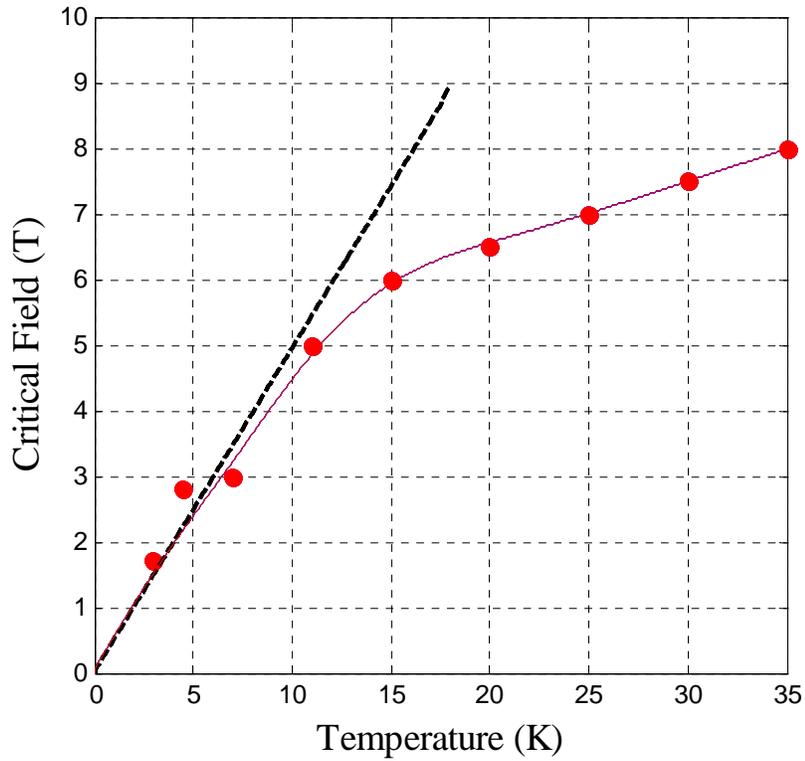

FIG. 9 Temperature dependence of the critical external magnetic field $H_C$ in ErMnO$_3$. Experimental points were determined from the 2D transmission intensity maps measured at different temperatures [similar to that is shown in Figure 8(c)]. Straight dotted line corresponds to $g_{Er}\mu_B H_C = k_B T$ calculated with $g_{Er} = 3$. Red curve guides the eye.